\documentclass[a4paper,11pt]{article}
\usepackage{pos}
\usepackage{amsmath}
\usepackage{graphicx}
\PassOptionsToPackage{normalem}{ulem}

\title{ Exclusive photoproduction of open heavy flavor meson pairs}

\author*{Marat Siddikov}

\affiliation{Departamento de Física, Universidad Técnica Federico Santa María,~~~~~~~\\
 y Centro Científico - Tecnológico de Valparaíso, Casilla 110-V, Valparaíso,
Chile}

\emailAdd{Marat.Siddikov@usm.cl}

\abstract{In this proceeding we argue that exclusive photoproduction
of $D$-meson pairs might be used as a complementary tool for studies of the generalized parton distributions of the target. We analyzed the photoproduction of the pseudoscalar-vector pairs with net zero electric charge (\emph{e.g.} $D^{\pm}D^{*\mp}$, $D^{0}\overline{D}^{*0}$, $D_{s}^{+}D_{s}^{*-}$) and found that it allows to study the chiral even GPDs in ERBL region. A unique feature of the suggested process is contribution of the gluon and just one of the light quark flavours.  We made numerical estimates in the kinematics of the future Electron Ion Collider and found that numerically the production cross-section is reasonably large for experimental studies, thus justifying its viability as a complementary probe of GPDs.}

\FullConference{25th International Spin Physics Symposium (SPIN 2023)\\
 24-29 September 2023\\
 Durham, NC, USA\\}

\begin{document}
\maketitle

\section{Introduction}

The Generalized Parton Distributions (GPDs) nowadays are one of
the widely used nonperturbative objects which describe the structure
of the hadronic target~\cite{Goeke:2001tz,Diehl:2003ny,Guidal:2013rya,Burkert:2022hjz},
and for this reason got into focus of various theoretical and experimental
studies. Due to the nonperturbative nature of GPDs, it is impossible to
find them directly from the first principles. Our current knowledge of these objects relies on lattice simulations~\cite{Egerer:2021ymv,Karpie:2021pap,Bhattacharya:2023nmv}, theoretical models based on additional assumptions (and thus requiring experimental confirmation), or outright phenomenological extractions from experimental data. The expected high-luminosity experiments at the future Electron Ion Collider motivated studies of various new channels which could provide better constraints on GPDs. For phenomenological studies of GPDs a special interest present the exclusive $2\to3$ processes~\cite{GPD2x3:9,GPD2x3:7,GPD2x3:6,GPD2x3:5,GPD2x3:4,GPD2x3:3,GPD2x3:2,GPD2x3:1,ElBeiyad:2010pji,Boussarie:2016qop,Pedrak:2020mfm}.
The factorization theorem for amplitudes of such processes has been
proven in the kinematics when all the produced hadrons are well-separated
kinematically~\cite{GPD2x3:10,GPD2x3:11}. Almost all these studies
focused on the production of light hadrons or photons by highly virtual
photons. In this proceeding we argue that it is possible to study
the GPDs using the photoproduction of heavier $D$-meson pairs. Previously,
the possibility to use $D$-mesons for studies of GPDs was discussed in~\cite{Pire:2015iza,Pire:2017lfj,Pire:2017yge,Pire:2021dad},
and the production of light meson pairs with large invariant was analyzed e.g. in~\cite{LehmannDronke:2000hlo,Clerbaux:2000hb}. The production of heavy meson pairs differs substantially from analogous producion of light mesons pairs, since the heavy quark mass breaks the conventional suppression based on twist counting, and thus eventually leads to new (independent) probes of the GPDs. In our study we consider production
of $D^{+}D^{*-}$,~$D^{0}\overline{D}^{*0}$ and $D_{s}^{+}D_{s}^{*-}$
meson pairs, whose cross-section is controlled by the chiral-even
GPDs. The choice of these final states allows to avoid contaminations by the
poorly known chiral odd transversity GPDs~\cite{ElBeiyad:2010pji,Boussarie:2016qop}
or photon-photon fusion mechanism suggested in~\cite{Luszczak:2011js}.

This proceeding is structured as follows. In the next Section~\ref{sec:Formalism}
we  define the kinematics and briefly introduce the framework used for evaluations of the cross-section of the process. In essence, we use for our analysis the conventional collinear factorization framework, treating the heavy quark mass as a hard scale (more details may be found in~\cite{Siddikov:2023qbd,Siddikov:2022bku}). In Section~\ref{sec:Numer}
we estimate numerically the cross-section using publicly available
parametrizations of the proton GPDs and $D$-meson distribution amplitudes in the kinematics of electron-proton collisions at the forthcoming Electron Ion Collider (EIC)~\cite{Accardi:2012qut,AbdulKhalek:2021gbh}.
\section{Exclusive photoproduction of meson pairs}

\label{sec:Formalism}
We will perform our evaluations in the photon-proton collision frame,
in which the light-cone decomposition of the photon momentum $q$
and proton momentum $p$ has a form

\begin{align}
q & =\left(-\frac{Q^{2}}{2q^{-}},\,q^{-},\,\,\boldsymbol{0}_{\perp}\right),\quad P=\left(P^{+},\,\frac{m_{N}^{2}}{2P^{+}},\,\,\boldsymbol{0}_{\perp}\right),\quad q^{-}=E_{\gamma}+\sqrt{E_{\gamma}^{2}+Q^{2}},\quad P^{+}=E_{p}+\sqrt{E_{p}^{2}-m_{N}^{2}}\label{eq:qPhoton-2}
\end{align}
where $Q^{2}=-q^{2}$ is the virtuality of the photon, and $E_{\gamma},\,E_{p}$
are energies of the photon and proton before collision. The 4-momenta
$p_{1},\,p_{2}$ of the final-state heavy $D$-mesons in this frame
have a light-cone decomposition
\begin{align}
p_{a} & =\left(\frac{M_{a}^{\perp}}{2}\,e^{-y_{a}}\,,\,M_{a}^{\perp}e^{y_{a}},\,\,\boldsymbol{p}_{a}^{\perp}\right),\quad M_{a}^{\perp}\equiv\sqrt{M_{a}^{2}+\left(\boldsymbol{p}_{a}^{\perp}\right)^{2}},\quad a=1,2,\label{eq:MesonLC-2}
\end{align}
where $y_{a}$ are the rapidities of produced mesons, and $\boldsymbol{p}_{a}^{\perp}$
are their transverse momenta. Since the production cross-section decreases
rapidly as a function of transverse momenta, in what follows we'll
focus on the kinematics of small momenta $p_{a}^{\perp}$. If we assume
that the invariant energy $W$ of the photon-proton collision is fixed,
then the onshellness condition for the 4-momentum of the recoil proton
$P'=P+\Delta$ translates into complicated constraints on possible
transverse momenta and rapidities of produced $D$-mesons. For this
reason, instead of conventional fixing of the invariant energy $W$
in electroproduction experiments it is easier to work with $D$-meson
momenta as unconstrained independent variables and fix energy of photon
from 4-momentum conservation. In this approach it is possible to rewrite
all kinematical variables in terms of rapidities and transverse momenta
of $D$-mesons. For example, the photon energy in this kinematics
might be approximated as $q^{-}\approx M_{1}^{\perp}\,e^{y_{1}}+M_{2}^{\perp}\,e^{y_{2}}$,
and the photoproduction cross-section is given by

\begin{equation}
d\bar{\sigma}_{\gamma p\to M_{1}M_{2}p}=\frac{dy_{1}dp_{1\perp}^{2}dy_{2}dp_{2\perp}^{2}d\phi\left|\mathcal{A}_{\gamma p\to M_{1}M_{2}p}\right|^{2}}{4\left(2\pi\right)^{4}W_{0}^{2}\sqrt{\left(W_{0}^{2}+Q^{2}-m_{N}^{2}\right)^{2}+4Q^{2}m_{N}^{2}}},\label{eq:Photo-1}
\end{equation}
In what follows we will assume that all the final state hadrons are
kinematically well-separated from each other. In this setup it is
possible to factorize the amplitude $\mathcal{A}_{\gamma p\to M_{1}M_{2}p}$
and express it as convolution of nonperturbative distributions of
final-state hadrons (target GPDs, distribution amplitudes of the $D$-mesons)
with perturbative partonic-level amplitudes, namely

\begin{align}
\sum_{{\rm spins}}\left|\mathcal{A}_{\gamma p\to M_{1}M_{2}p}^{(\mathfrak{a})}\right|^{2} & =\frac{1}{\left(2-x_{B}\right)^{2}}\left[4\left(1-x_{B}\right)\left(\mathcal{H}_{\mathfrak{a}}\mathcal{H}_{\mathfrak{a}}^{*}+\tilde{\mathcal{H}}_{\mathfrak{a}}\tilde{\mathcal{H}}_{\mathfrak{a}}^{*}\right)-x_{B}^{2}\left(\mathcal{H}_{\mathfrak{a}}\mathcal{E}_{\mathfrak{a}}^{*}+\mathcal{E}_{\mathfrak{a}}\mathcal{H}_{\mathfrak{a}}^{*}+\tilde{\mathcal{H}}_{\mathfrak{a}}\tilde{\mathcal{E}}_{\mathfrak{a}}^{*}+\tilde{\mathcal{E}}_{\mathfrak{a}}\tilde{\mathcal{H}}_{\mathfrak{a}}^{*}\right)\right.\label{eq:AmpSq}\\
 & \qquad\left.-\left(x_{B}^{2}+\left(2-x_{B}\right)^{2}\frac{t}{4m_{N}^{2}}\right)\mathcal{E}_{\mathfrak{a}}\mathcal{E}_{\mathfrak{a}}^{*}-x_{B}^{2}\frac{t}{4m_{N}^{2}}\tilde{\mathcal{E}}_{\mathfrak{a}}\tilde{\mathcal{E}}_{\mathfrak{a}}^{*}\right],\qquad\mathfrak{a}=L,T\nonumber
\end{align}
where the subscript index $\mathfrak{a}$ differentiates the longitudinal
and transverse polarization of the incident photon, and, motivated
by the earlier analyses of DVCS and DVMP~\cite{Belitsky:2001ns,Belitsky:2005qn},
we defined the double meson form factors as

\begin{align}
 & \left.\begin{array}{c}
\mathcal{H}_{\mathfrak{a}}\left(\xi,\,\Delta y,t\right)\\
\mathcal{E}_{\mathfrak{a}}\left(\xi,\,\Delta y,t\right)
\end{array}\right\} =\sum_{\kappa=q,g}\int_{-1}^{1}dx\prod_{n=1}^{2}\left(\int_{0}^{1}dz_{n}\varphi_{D_{n}}\left(z_{n}\right)\right)C_{\mathfrak{a}}^{(\kappa)}\left(x,\,\xi,\,\Delta y,z_{1},\,z_{2}\right)\left\{ \begin{array}{c}
H_{\kappa}\left(x,\xi,t\right)\\
E_{\kappa}\left(x,\xi,t\right)
\end{array}\right.,\label{eq:Ha}\\
 & \left.\begin{array}{c}
\tilde{\mathcal{H}}_{\mathfrak{a}}\left(\xi,\,\Delta y,t\right)\\
\tilde{\mathcal{E}}_{\mathfrak{a}}\left(\xi,\,\Delta y,t\right)
\end{array}\right\} =\sum_{\kappa=q,g}\int_{-1}^{1}dx \prod_{n=1}^{2}\left(\int_{0}^{1}dz_{n}\varphi_{D_{n}}\left(z_{n}\right)\right)\,\tilde{C}_{\mathfrak{a}}^{(\kappa)}\left(x,\,\xi,\,\Delta y,t\right)\times\left\{ \begin{array}{c}
\tilde{H}_{\kappa}\left(x,\xi,t\right)\\
\tilde{E}_{\kappa}\left(x,\xi,t\right)
\end{array}\right..\label{eq:ETildeA}
\end{align}
The dummy integration variables $z_{1},\,z_{2}$ correspond to the
light-cone fractions of the total momentum carried by $c$-quarks in
$D$-mesons. Since we consider that the final-state $D$-mesons are well-separated from each other, the Fock state of the 2-meson system is a direct
product of Fock states of individual $D$-mesons. In the heavy quark
mass limit the Fock state of each $D$-meson is dominated by the 2-quark
component described by $D$-meson distribution amplitudes~$\varphi_{D}\left(z\right)$~\cite{Zuo:2006re}.
The partonic amplitudes $C_{\mathfrak{a}}^{(\kappa)},\,\tilde{C}_{\mathfrak{a}}^{(\kappa)}$
can be evaluated perturbatively, taking into account all the quark
and gluon diagrams which might contribute to the final state. Due
to space limitations here we omit explicit expressions for $C_{\mathfrak{a}}^{(\kappa)},\,\tilde{C}_{\mathfrak{a}}^{(\kappa)}$;
the reader may find them in~\cite{Siddikov:2023qbd}. We may see
that in~(\ref{eq:Ha},\ref{eq:ETildeA}) the GPDs contribute convoluted
with effective (integrated) coefficient function
\begin{align}
C_{{\rm int}}^{(\kappa)}\left(x,\,\xi,\,\Delta y\right) & \equiv\int_{0}^{1}dz_{1}\int_{0}^{1}dz_{2}\,\,\varphi_{D_{1}}\left(z_{1}\right)\varphi_{D_{2}}\left(z_{2}\right)C_{T}^{(\kappa)}\left(x,\,\xi,\,\Delta y,\,z_{1},\,z_{2}\right).\label{eq:Cintq}
\end{align}
Due to convolution with relatively broad $D$-meson distribution amplitudes
in~(\ref{eq:Cintq}), the functions $C_{{\rm int}}^{(\kappa)}$ do
not have any singularities and are strongly concentrated in the region
$|x|\le\xi$ (the so-called ERBL region), which implies that the cross-section
of the process is mostly sensitive to the behavior of GPDs in that
domain.

\section{Numerical results}

\label{sec:Numer} For definiteness, we will use for our estimates
the Kroll-Goloskokov parametrization of the GPDs~\cite{Goloskokov:2011rd,Goloskokov:2013mba},
fixing the factorization scale as $\mu_{F}=\mu_{R}=4\,{\rm GeV}\approx2M_{D}$.
Due to space limitations here we will consider only the production
of $D_{0}\,\overline{D_{0}}$ meson pairs (an interested reader may
find predictions for other mesons in~\cite{Siddikov:2023qbd}). In
general, the cross-sections for all meson pairs have qualitatively
the same dependence on kinematic variables, though the absolute values
can differ by up to an order of magnitude. The $Q^{2}$-dependence
of the photoproduction cross-section is controlled by a relatively
large invariant mass of the produced pair $\left(M_{1}+M_{2}\right)$,
and in the photoproduction region which analyzed here this dependence is very weak.

\begin{figure}
\includegraphics[width=5cm]{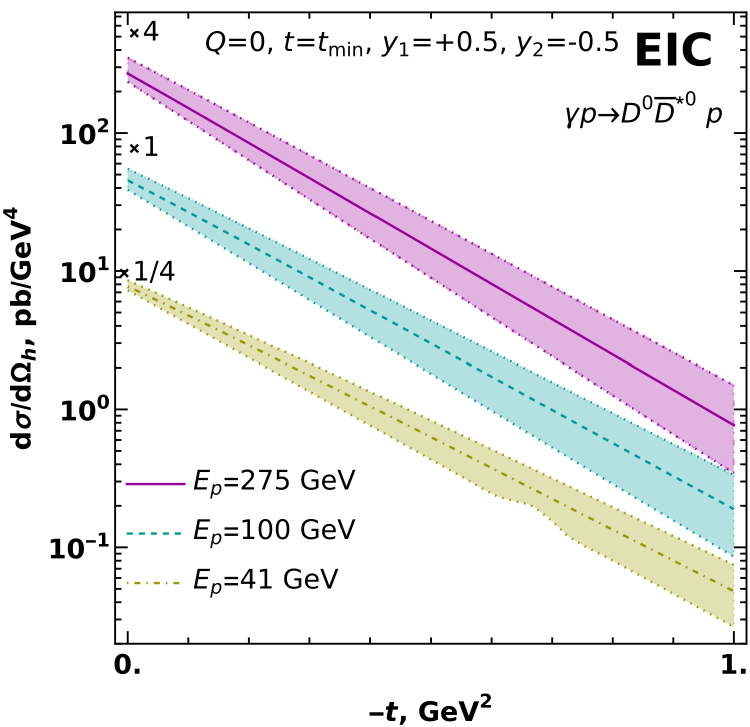}\includegraphics[width=5cm]{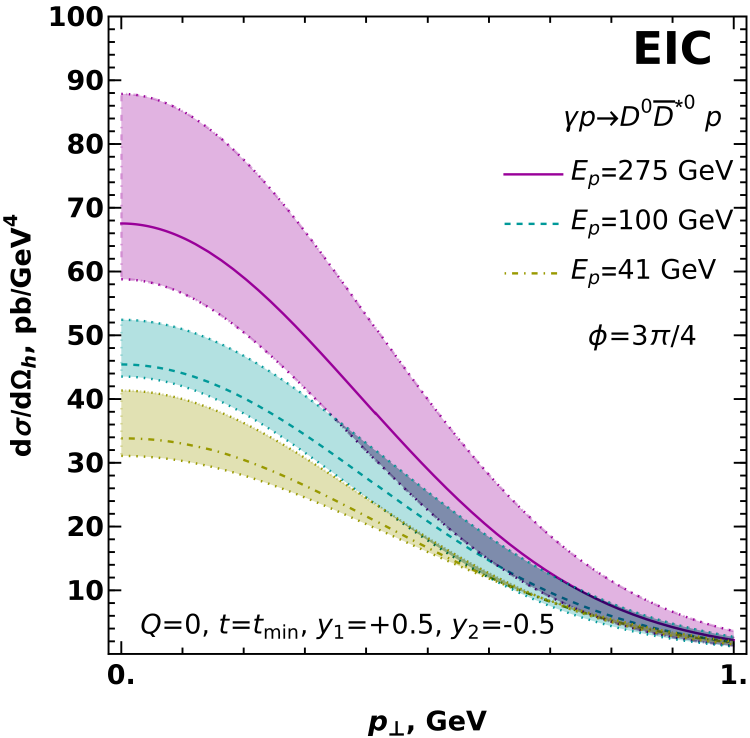}\includegraphics[width=5cm]{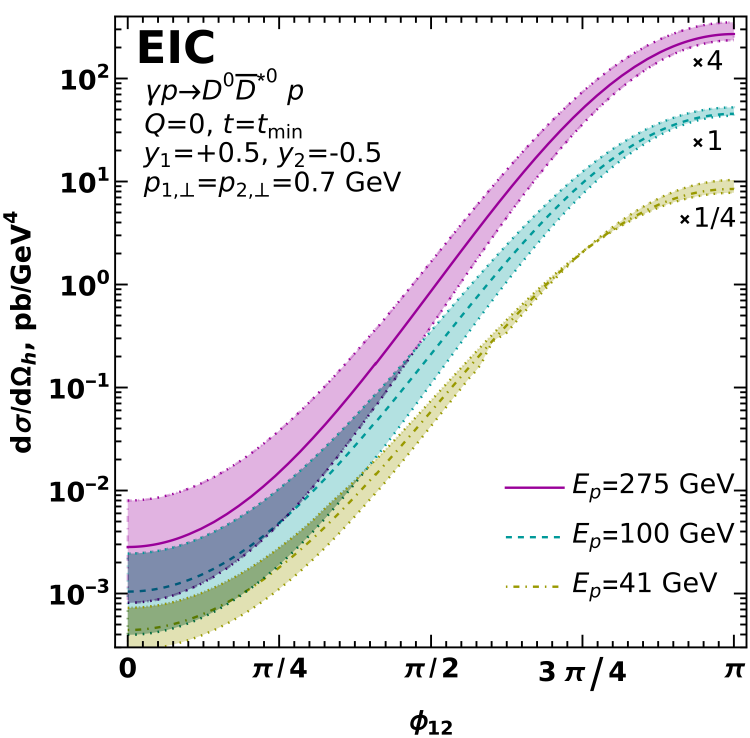}

\caption{\label{fig:tDep-1} The dependence of the production cross-sections
on the invariant momentum transfer $t$ (left), on transverse momentum
$p_{T}$ (central), and on the azimuthal angle $\phi_{12}$ (right).
The width of the colored band in each plot illustrates the uncertainty
due to choice of the factorization scale $\mu_{F}\in\left(0.5,2\right)\times2m_{D}$.
For the sake of legibility in the left and right plots we multiplied
the cross-sections for $E_{p}={\rm 275}\,{\rm GeV}$ and $E_{p}=41\,{\rm GeV}$
by numerical factors $\times4$ and $\times1/4$ respectively.}
\end{figure}

The Figure~(\ref{fig:tDep-1}) illustrates the dependence of the
cross-section~(\ref{eq:Photo-1}) on the invariant momentum transfer
$t$ , and the associated distributions of the $D$-meson pairs on
transverse momenta and azimuthal angle between them in the photon-proton frame.
Since in the collinear factorization framework the transverse momenta
are disregarded in the coefficient function, the $t$-dependence originates
entirely from the implemented GPDs. The phenomenological studies show
that this dependence should be exponentially suppressed as a function
of $|t|$, which implies that in the photon-proton frame the $D$-meson
pairs are produced predominantly in the back-to-back kinematics, with
minimal momentum transfer $t$ to the target.

The cross-section grows mildly as a function of average rapidity $Y=(y_{1}+y_{2})/2$
of the produced quarkonia, yet decreases rapidly as a function of
rapidity difference $\Delta y=\left|y_{1}-y_{2}\right|$. This behavior
might be understood if we recall that the increase of $Y$ at fixed
$\Delta y$ increases the invariant energy $W^{2}$, decreases $x_{B},\,\xi$
and thus leads to growth of the partonic GPDs. On the other hand, the
increase of rapidity difference $\Delta y$ at fixed $Y$, due to onshellness
constraints for recoil proton, leads to increase of the momentum transfer
$|t|$ and corresponding suppression of the partonic GPDs (detailed
plots for rapidity dependence of different mesons may be found in~\cite{Siddikov:2023qbd}).

To summarize, we believe that the exclusive
photoproduction of the pseudoscalar-vector $D$-meson pairs ($D^{+}D^{*-},$~$D^{0}\bar{D}^{*0}$ and $D_{s}^{+}D_{s}^{*-}$) can be used as complementary
probe of chiral-even GPDs of gluon and one of the light quark flavors in the so-called ERBL region $|x|<\xi$. Numerically, the cross-sections of the proposed channels are comparable to the cross-sections of other $2\to 3$ processes suggested in the literature~\cite{GPD2x3:9,GPD2x3:7,GPD2x3:6,GPD2x3:5,GPD2x3:4,GPD2x3:3,GPD2x3:2,GPD2x3:1,ElBeiyad:2010pji,Boussarie:2016qop}.

\section*{Acknowldgements}

We thank our colleagues at UTFSM university for encouraging discussions.
This research was partially supported by Proyecto ANID PIA/APOYO AFB220004
(Chile) and Fondecyt (Chile) grants 1220242. \textquotedbl Powered@NLHPC:
This research was partially supported by the supercomputing infrastructure
of the NLHPC (ECM-02)\textquotedbl .

\end{document}